\documentclass[pra,aps,superscriptaddress,twocolumn,showpacs]{revtex4}
\usepackage{graphicx}
\usepackage{dcolumn}
\usepackage{bm}
\usepackage{amssymb}
\usepackage{amsmath}
\usepackage{amsthm}
\usepackage{chemarrow}
\usepackage{multirow}
\usepackage{color}
\usepackage{bigstrut}
\usepackage{epstopdf}
\begin{document}
\title{Preparing Greenberger-Horne-Zeilinger and $W$ states on a long-range Ising spin model by global controls}
\author{Jiahui Chen}\thanks{Present address: Institute for Quantum Computing and
Department of Physics and Astronomy, University of Waterloo, Waterloo, Ontario N2L 3G1, Canada}
\affiliation{Hefei National Laboratory for Physical Sciences at Microscale and Department of Modern Physics,
University of Science and Technology of China, Hefei, Anhui 230026, China}
\author{Hui Zhou}
\affiliation{Department of Physics, Shaanxi University of Science and Technology, Xi'an 710021, China}
\author{Changkui Duan}\email{ckduan@ustc.edu.cn}
\affiliation{Hefei National Laboratory for Physical Sciences at Microscale and Department of Modern Physics,
University of Science and Technology of China, Hefei, Anhui 230026, China}
\author{Xinhua Peng}\email{xhpeng@ustc.edu.cn}
\affiliation{Hefei National Laboratory for Physical Sciences at Microscale and Department of Modern Physics,
University of Science and Technology of China, Hefei, Anhui 230026, China}
\affiliation{College of Physics and Electronic Science, Hubei Normal University, Huangshi, Hubei 435002, China}

\date{\today}
\begin{abstract}
Entanglement, a unique quantum resource with no classical counterpart, remains at the heart of quantum information. The Greenberger-Horne-Zeilinger (GHZ) and $W$ states are two inequivalent classes of multipartite entangled states which can not be transformed into each other by means of local operations and classic communication. In this paper, we present the methods to prepare the GHZ and $W$ states via global controls on a long-range Ising spin model. For the GHZ state, general solutions are analytically obtained for an arbitrary-size spin system, while for the $W$ state, we find a standard way to prepare the $W$ state that is analytically illustrated in three- and four-spin systems and numerically demonstrated for larger-size systems. The number of parameters required in the numerical search increases only linearly with the size of the system.
\end{abstract}
\maketitle

\section{\label{sec:level1}Introduction}

Entanglement is one of the most intriguing features of quantum physics, which remains at the heart of applications such as quantum computation \cite{qu-com}, quantum teleportation \cite{qu-tele}, and quantum cryptography \cite{CrApp, CrPhoton}. Meanwhile, it plays a crucial role in a variety of phenomena, e.g., the fractional quantum Hall effect \cite{qu-ha} and quantum phase transitions \cite{qu-tr}. Therefore, preparation of the entangled states is of significance in many-particle physics. It is well-known that there are two important and different types of entangled states, which cannot be transformed into each other under local operations and classical communication \cite{LOCC}, i.e., the Greenberger-Horne-Zeilinger (GHZ) state \cite{GHZstate}
\begin{equation}
\left | {\rm GHZ} \right \rangle =\frac{1}{\sqrt{2}}\left ( \left | 000\cdots0 \right \rangle+\left | 111\cdots1 \right \rangle \right )
\end{equation}
and the $W$ state \cite{Wstate}
\begin{equation}
\begin{aligned}
\left | W \right \rangle = \frac{1}{\sqrt{n}}\left ( { \left | 100\cdots0 \right \rangle+\left | 010\cdots0 \right \rangle } \right. \\
\left. {  + \left | 001\cdots0 \right \rangle +\cdots + \left | 000\cdots1 \right \rangle} \right ).\\
\label{Wstate}
\end{aligned}
\end{equation}

The study of these two states has attracted much interest
\cite{Property}, and different methods to prepare these two states have
been proposed recently, e.g. via dissipative preparation of
entangled states \cite{11,12}.

One of the most commonly used methods to prepare entangled states is via quantum circuits, which has been implemented in experiments on various kinds of quantum systems, such as nuclear magnetic resonance (NMR) \cite{QC-NMR}, ion traps \cite{QC-ion} and cavity QED \cite{QC-QED}. However, this method usually requires individual addressability of qubits, which makes the experiment difficult in large systems with many qubits. To overcome this problem, one of the practical ways is to employ global controls that act on all the spins.

The Ising spin model is one of the most ubiquitous in many physical systems such as optical lattices \cite{jack}, NMR systems \cite{Vandersypen}, ion traps \cite{porras} and polar molecules \cite{Micheli}. It also plays an important role in both condensed-matter physics \cite{ising-cmp} and quantum information theory \cite{ising-qit}. Recently, the study of global control methods to generate entangled states has attracted a lot of attention in the Ising-type spin-spin interaction systems \cite{global control GHZ, cluster,ben,gao}. The global control method relaxes the demanding experimental requirement to address and operate a single spin.

In this paper, we study how to prepare GHZ and $W$ states on a long-range Ising spin model via global controls. The rest of this paper is organized as follows. In Sec. II we introduce the studied spin model under global controls and our quantum control problem. In Sec. III we present the general solutions to obtain the GHZ state. In Sec. IV, we establish a standard way to generate the $W$ state. Finally, a brief summary with a discussion is presented in Sec. V.

\section{\label{sec:level2} long-range Ising spin model and control problem}

An $n$-spin long-range Ising spin model has the following Hamiltonian:
 \begin{equation}
H_{\text{Ising}}=H_{zz} = \sum_{k<m}^{n}\sigma _{k}^{z}\sigma _{m}^{z} ,
\label{HIsing}
\end{equation}
where $\sigma_k^{\alpha}\left(\alpha=x,y,z\right)$ are the spin-1/2 Pauli matrices acting on the $k$th qubit.  Here, we adopt the following two available global controls
\begin{equation}
H_{x}=\sum_{k=1}^{n}\sigma _{k}^{x} ,  \, \,  \, \, \,  H_{y}=\sum_{k=1}^{n}\sigma _{k}^{y}
\label{Hcontrol},
\end{equation}
i.e., applying transverse magnetic fields on all of the spins along the $x$ or $y$ direction.

The total time-dependent Hamiltonian is
\begin{equation}
H \left ( t \right )=H_{zz}+f\left ( t \right )H_{x}+g\left ( t \right )H_{y},
\label{e.gc}
\end{equation}
where $f(t)$ and $g(t)$ are time-dependent functions. We assume here that the global fields are strong enough that the pulses can be regarded as the form of a $\delta$ function. This condition is easily satisfied in some physical systems such as NMR. This implies that the propagator can be regarded as a product of the time evolution operators under Hamiltonians in Eqs. \eqref{HIsing} and \eqref{Hcontrol}:
\begin{equation}
U(M, \tau_m, \beta _m^x,\beta _m^y)  = \prod\limits_{m =1}^{M} {{e^{ - i{H_{zz}}{\tau _m}}}{e^{ - i\beta _m^x {H_x }}}{e^{ - i\beta _m^y {H_y }}}}\label{evolution}.
\end{equation}
Our task is to find a pulse sequence with a set of suitable parameters $\{M, \tau_m, \beta _m^x,\beta _m^y \}$ to maximize the fidelity
\begin{equation}
F\left[U\right]=\left|\left\langle T\right|U(M, \tau_m, \beta _m^x,\beta _m^y) \left|00\cdots0\right\rangle\right|.\label{fidelity}
\end{equation}
Here, $|T\rangle = | \text{GHZ} \rangle$ or $| W \rangle$. In the following sections, let
\begin{equation}
\begin{split}
&ZZ(\tau)=e^{-i \tau H_{zz}},\\
&X(\beta)=e^{-i\beta H_x},\\
&Y(\beta)=e^{-i \beta H_y}.
\end{split}
\end{equation}

%In experiments, time parameters can never be negative. On the other hand, each time evolution operator considered belongs to a compact Lie group, it obeys the periodicity. Hence, for convenience, we allow the parameters for global controls to be negative since they do not essentially cost any time in our consideration while keep the positivity for the time parameter for the free evolution.

\section{GHZ state}
In this section, we discuss the cases in odd- and even-spin systems, respectively.

\subsection{Odd-spin systems}
Inspired by the solution in three-spin system \cite{gao}, we found the general sequence for preparing the GHZ state in odd-spin systems as
\begin{equation}
Y(\pi/4)-ZZ(\pi/4)-X(\pi/4).
\end{equation}
We will prove this in the following.
A $Y(\pi/4)$ pulse applied to the initial state $|0\cdots00\rangle$ yields a uniform superposition:
\begin{equation}
Y\left(\frac{\pi}{4}\right)|0\cdots00\rangle=\frac{1}{\sqrt{2^n}}\sum\limits_{k = 0}^{2^n-1}{|k\rangle}.
\label{initialp}
\end{equation}
Here, $k$ is a binary number indicating the index of the basis vector. Any vector from the computational basis is an eigenvector of the interaction Hamiltonian:
\begin{equation}
H_{\text{Ising}}|k\rangle=\left\{\frac{n(n-1)}{2}-2s(k)\left[n-s(k)\right]\right\}|k\rangle,
\end{equation}
where $s(k)$ is number of spin-up spins in the state $|k\rangle$. Therefore,
\begin{equation}
\begin{split}
& ZZ\left(\frac{\pi}{4}\right)Y\left(\frac{\pi}{4}\right)|0\cdots00\rangle\\
& = \frac{1}{\sqrt{2^n}}\sum\limits_{k = 0}^{2^n-1} e^{-i\left\{\frac{n(n-1)}{2}-2s(k)\left[n-s(k)\right]\right\}\frac{\pi}{4}}|k\rangle\\
&= e^{-i\frac{n(n-1)\pi}{8}} \frac{1}{\sqrt{2^n}}\sum\limits_{k = 0}^{2^n-1}i^{s(k)\left[n-s(k)\right]}|k\rangle.
\label{zzco}
\end{split}
\end{equation}
Applying $X(-\pi/4)$ on the GHZ state, we obtain
\begin{equation}
\begin{split}
&X\left(-\frac{\pi}{4}\right)\left|\text{GHZ}\right\rangle\\
&=\frac{1}{\sqrt{2}^{n+1}}\left[(|0\rangle+i|1\rangle)^{\otimes n}+(i|0\rangle+|1\rangle)^{\otimes n}\right]\\
&=\frac{1}{\sqrt{2}^{n+1}}\sum\limits_{k = 0}^{2^n-1}\left(i^{s(k)}+i^{n-s(k)}\right)|k\rangle.
\end{split}
\label{GHZh}
\end{equation}
Except for the global phase, the ratio of the coefficient of the state $|k\rangle$ in Eq. \eqref{GHZh} to the coefficient of Eq. \eqref{zzco} is a function of $s(k)$:
\begin{equation}
\begin{split}
f(s(k))&=\left(i^{s(k)}+i^{n-s(k)}\right)/i^{s(k)[n-s(k)]}\\
&=i^{s(k)-s(k)[n-s(k)]}+i^{n-s(k)-s(k)[n-s(k)]}\\
&=i^{s(k)[1-n+s(k)]}+i^{[1-s(k)][n-s(k)]}.
\end{split}
\end{equation}
Hence,
\begin{equation}
f(s(k)+1)=i^{[s(k)+1][2-n+s(k)]}+i^{s(k)[1-n+s(k)]}.
\end{equation}
Note that $i^{[1-s(k)][n-s(k)]}/i^{[s(k)+1][2-n+s(k)]}=i^{2[n-2s(k)-1]}=1$ when $n$ is odd. Therefore, we have
\begin{equation}
f(s(k))=f(s(k)+1)
\label{dengdeng},
\end{equation}
which implies that $f(s(k))$ is independent of $s(k)$ and thus the state in Eq. \eqref{GHZh} is equal to that of Eq. \eqref{zzco} up to an overall phase. Therefore the sequence $Y(\pi/4)-ZZ(\pi/4)-X(\pi/4)$ creates the GHZ state in odd systems.

\subsection{Even-spin systems}
For even systems, preparing GHZ states can be regarded as creating the total spin coherence between the state $|00\cdots0\rangle$ and $|11\cdots1\rangle$ with a specific phase difference \cite{slichter}. The $ZZ$ coupling alone is zero coherence, which cannot change the coherence order of a state. However, the $ZZ$ coupling can be transformed into $ H_{xx}=\sum_{k<m}^{n}\sigma _{k}^{x}\sigma _{m}^{x}=e^{-iH_y\pi/4}H_{zz}e^{iH_y\pi/4}$, which is a sum of zero-coherence and double-coherence terms. With this Hamiltonian, it is more convenient to expand the initial state as a sum of tensor products of $\sigma^x$'s eigenstates $\left\{|+\rangle,|-\rangle\right\}$. Let $|+\rangle=|0\rangle_x$ and $|-\rangle=|1\rangle_x$, and then $|k\rangle_x$ is the binary representation in the $\sigma^x$'s basis. The initial state can be expanded as
\begin{equation}
|00\cdots0\rangle=\frac{1}{\sqrt{2^n}}\left(|+\rangle+|-\rangle\right)^{\otimes n}=\frac{1}{\sqrt{2^n}}\sum\limits_{k = 0}^{2^n-1}{|k\rangle_x}
\label{iniexp}.
\end{equation}
Like for Eq. \eqref{zzco}, we have
\begin{equation}
XX\left(\frac{\pi}{4}\right)|k\rangle_x=e^{-i\frac{n(n-1)\pi}{8}}i^{s_x(k)[n-s_x(k)]}|k\rangle_x
\label{XXco},
\end{equation}
where $s_x(k)$ is the number of plus signs in state $|k\rangle_x$ and $XX\left(t\right)=e^{-iH_{xx}t}$ is the evolution operator under the Hamiltonian $H_{xx}$. With the same method used in Eq. \eqref{dengdeng}, it can be proved that $i^{s_x(k)[n-s_x(k)]}/\left[1+(-1)^{n-s_x(k)}i^{n+1}\right]$ is independent of $s_x(k)$ when $n$ is even. Combining this result with Eqs. \eqref{iniexp} and \eqref{XXco}, regardless of the global phase, we have
\begin{equation}
\begin{split}
&XX\left(\frac{\pi}{4}\right)|00\cdots0\rangle\\
&=\frac{1}{\sqrt{2}^{n+1}}\sum\limits_{k = 0}^{2^n-1}{\left[1+(-1)^{n-s_x(k)}i^{n+1}\right]|k\rangle_x}\\
&=\frac{1}{\sqrt2}\left(|00\cdots0\rangle+i^{n+1}|11\cdots1\rangle\right).
\end{split}
\end{equation}
This is the GHZ-type state. In order to obtain the exact state $| \text{GHZ} \rangle$ in Eq. (1), one needs to further apply a global $Z$ rotation:
\begin{equation}
Z\left(-\frac{(n+1)\pi}{4n}\right)=Y\left(\pi/4\right)-X\left(-\frac{(n+1)\pi}{4n}\right)-Y\left(-\pi/4\right).
\end{equation}
In addition, as we discussed, the operator $XX(t)$ can be implemented as $Y(\pi/4)-ZZ(t)-Y(-\pi/4)$. Accordingly the sequence to prepare the GHZ state in even-spin systems is
\begin{equation}
Y\left(\pi/4\right)-ZZ\left(\pi/4\right)-X\left(-\frac{(n+1)\pi}{4n}\right)-Y\left(-\pi/4\right).
\end{equation}

\section{\label{sec:level2} $W$ state}

In this section, we study how to generate the $W$ state.
%Since the dimension of the Hilbert space exponentially increases with the number of qubits $n$, it is difficult to
%find the solution to generate the $W$ state. However, it is still possible to simplify the calculation with help of the symmetry of the system.
Since the Hamiltonian \eqref{HIsing} and the global control Hamiltonians
\eqref{Hcontrol} are symmetric with respect to permutation of the qubits, they can be block diagonalized under the following symmetry-adapted basis:
\begin{equation}
\left|\varphi _{m}\right\rangle=\frac{1}{\sqrt{C^{m-1}_n}}\sum\limits_{s(k)=m-1}|k\rangle.
\label{basis}
\end{equation}
Here, $C^{m}_n$ is the number of $m$ combinations over $n$ elements; the sum is taken over all the computational basis states $|k\rangle$ with $s(k)$ spin-up spins, $m=1,\cdots,n+1$. The initial state $|0\cdots00\rangle$ and the target state $|W\rangle$ are within subspace spanned by basis \eqref{basis}, implying that the calculation can be analyzed in this subspace.
Moreover, $H_{zz}$ and $H_x$ commute with the ``$X$-parity" operator $X = \prod_{i=1}^n \sigma_i^x$; therefore, the representations of $H_{zz}$ and $H_x$ in the symmetry-adapted basis can be further simultaneously block diagonalized in the two eigenspaces of $X$ with eigenvalues $\pm1$:
\begin{equation}
\mathcal{X}_\pm:   \left|x^\pm_k\right\rangle=c\left(\left|\varphi_k\right\rangle\pm\left|\varphi_{n+2-k}\right\rangle\right),
\label{xbasis}
\end{equation}
where $c$ is the normalizing constant and $k=1,2,\cdots,\left\lfloor n/2\right\rfloor+1$ ($\left\lfloor x\right\rfloor$ is the largest integer less than or equal to $x$).
Likewise, the representations of $H_{zz}$ and $H_y$ in the symmetry-adapted basis can be block-diagonalized under the eigenstates of the ``$Y$-parity" operator $Y = \prod_{i=1}^n \sigma_i^y$:
\begin{equation}
\mathcal{Y}_\pm:  \left|y^\pm_k\right\rangle=c\left[\left|\varphi_k\right\rangle\pm i^n\left(-1\right)^{k-1}\left|\varphi_{n+2-k}\right\rangle\right],
\label{ybasis}
\end{equation}
where $c$ is the normalizing constant and $k=1,2,\cdots,\left\lfloor n/2\right\rfloor+1$.

\begin{figure}[htbp]
\centering
\includegraphics[width=0.42\textwidth]{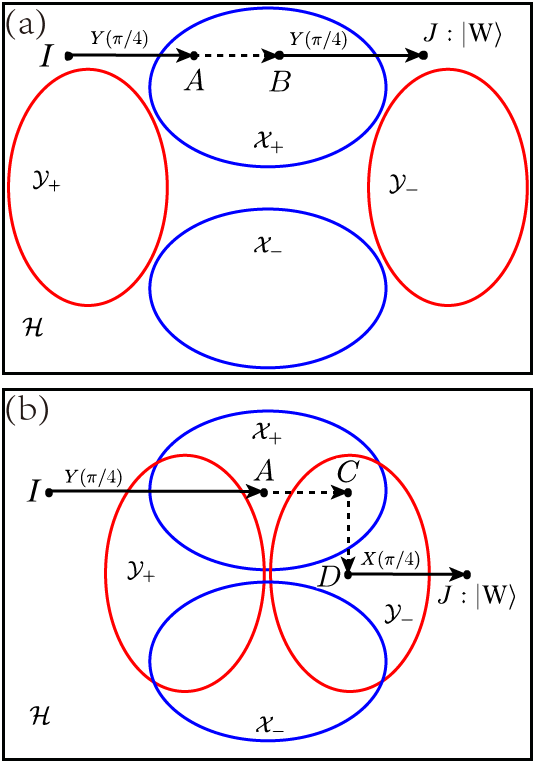} \caption{Schematic of preparing the $W$ state for (a) the odd-spin system and (b) the even-spin system. Black squares represent all quantum states from the total Hilbert space $\mathcal H$. The blue ovals and the red ovals denote the subsets of quantum states from $\mathcal{X}_+, \mathcal{X}_-$ and  $\mathcal{Y}_+, \mathcal{Y}_-$, respectively. Solid arrows represent single pulses; dashed arrows represent corresponding evolutions in subspaces. Black points represent quantum states; $A$, $B$, $C$, and $D$ are intermediate states in corresponding subspaces. The initial state $|00\cdots0\rangle$ (point $I$) and the $W$ state (point $J$) are  not in any subspace. We note here that when $n=2$, the $|W\rangle$ state is within the subspace $\mathcal{X}_+$, thus the evolution from the point $C$ to the point $D$ is not necessary in this case.
}\label{scheme}
\end{figure}

Using different operations ($X, Y$, and $ZZ$ operations), we can shuttle between these subspaces. According to Eqs. \eqref{basis}, \eqref{xbasis}, and \eqref{ybasis}, the following properties can be obtained when $n\geqslant2$ (the corresponding transformations are represented in Fig. \ref{scheme}): (1) When $n$ is odd [Fig. \ref{scheme} (a)]

(i) $Y(\pi/4)$ rotates $\left|00\cdots0\right\rangle$ (point $I$) into the subspace $\mathcal{X}_+$ (point $A$),
\begin{equation}
Y\left(\frac{\pi}{4}\right)\left|00 \cdots 0\right\rangle
=\sum\limits_{k = 0}^{\frac{n - 1}{2}} {\sqrt {\frac{{C_n^k}}{{{2^{n - 1}}}}} } \left| {x_{k + 1}^ + } \right\rangle
\label{initial2};
\end{equation}

and (ii) $Y(-\pi/4)$ rotate the $W$ state (point $J$) into the subspace $\mathcal{X}_+$ (point $B$),
\begin{equation}
Y\left(-\frac{\pi}{4}\right)\left|W\right\rangle=\sum\limits_{k = 0}^{\frac{n-1}{2}} {\sqrt{\frac{C^k_n}{2^{n-1}n} }\left(-1\right)^k\left(n-2k\right)\left|x^+_{k+1}\right\rangle}.\label{Wodd}
\end{equation}
(2) When $n$ is even. [Fig. \ref{scheme} (b)],
(i) $Y(\pi/4)$ rotates $\left|00\cdots0\right\rangle$ (point $I$) into the subspace $\mathcal{X}_+$ (point $A$),
\begin{equation}
Y\left(\frac{\pi}{4}\right)\left|00 \cdots 0\right\rangle\\=\sum\limits_{k = 0}^{\frac{n}{2} - 1} {\sqrt {\frac{{C_n^k}}{{{2^{n - 1}}}}} } \left| {x_{k + 1}^ + } \right\rangle  + \sqrt {\frac{{C_n^{n/2}}}{{{2^n}}}} \left| {x_{n/2 + 1}^ + } \right\rangle,
\label{initial}
\end{equation}

and (ii) $X(-\pi/4)$ rotates the $W$ state (point $J$) into the subspace $\mathcal{Y}_-$ (point $D$),
\begin{equation}
X \left(-\frac{\pi}{4}\right)\left|W\right\rangle=
  \sum\limits_{k = 0}^{\frac{n}{2}-1}{\sqrt{\frac{C^k_n}{2^{n-1}n} }\left[i^{k-1}\left(2k-n\right)\right]\left|y^-_{k+1}\right\rangle}.\label{Weven}
\end{equation}\newline

%If we rotate the initial state and the target states into the same invariant subspace ($\mathcal X_\pm$ or $\mathcal Y_\pm$) via global controls so that we can search for a solution within the subspace.Since subspaces $\mathcal{X}_\pm$ partially overlap the subspaces $\mathcal{Y}_\pm$ when $n$ is even, if the target state can not be rotated into the same subspace with the initial state ($\mathcal X_+$), we can rotate it into a different subspace ($\mathcal Y_-$ or $\mathcal Y_+$) that overlaps the subspace $\mathcal X_+$.

We also note here that the subspaces $\mathcal X_\pm$ overlap the subspaces $\mathcal Y_\pm$ when $n$ is even. Utilizing these properties, a routine way to generate the $W$ state can be summarized as:
\begin{equation}
Y\left(\frac{\pi}{4} \right) - U_x-
                            \begin{cases}U_y-X\left(\frac{\pi}{4}\right)&n\text{ is even},\\
                            Y\left(\frac{\pi}{4}\right)&n\text{ is odd},\end{cases}
\label{steps}
\end{equation}
where $U_x = \prod_i {ZZ(\theta_{zzi}) X(\theta_{xi})} $ and $U_y = \prod_i{ZZ(\theta_{zzi}) Y(\theta_{yi})} $.
The number of operations needed here is uniformly bounded \cite{shandeluo}. Since the dimension of the subspace increases linearly with the size $n$ of the system, we expect that the number of operations required also has a linear dependence on $n$. In addition, the system is pure state controllable in the subspaces $\mathcal X_\pm$ and $\mathcal Y_\pm$, respectively; a solution through the decomposition \eqref{steps} thus always exists. The proof is included in the Appendix. Therefore, a solution with high fidelity can always be found by setting the error threshold sufficiently small in principle.

As examples, we show how to prepare the $W$ state on three- and four-spin Ising models. Here, we denote the matrix representations of different Hamiltonians in different subspaces by $H_\alpha^{\Xi}$, where $\alpha=x$, $y$ or $zz$ and $\Xi=\mathcal{X}_\pm$ or $\mathcal{Y}_\pm$ represents the corresponding subspaces. We use the right-hand side of Eqs. \eqref{initial2}  and \eqref{initial} as the initial states, and the
right-hand side of Eqs. \eqref{Wodd}  and  \eqref{Weven} as the target states for odd and even $n$, respectively.

\begin{figure}
\includegraphics[width=0.5\textwidth]{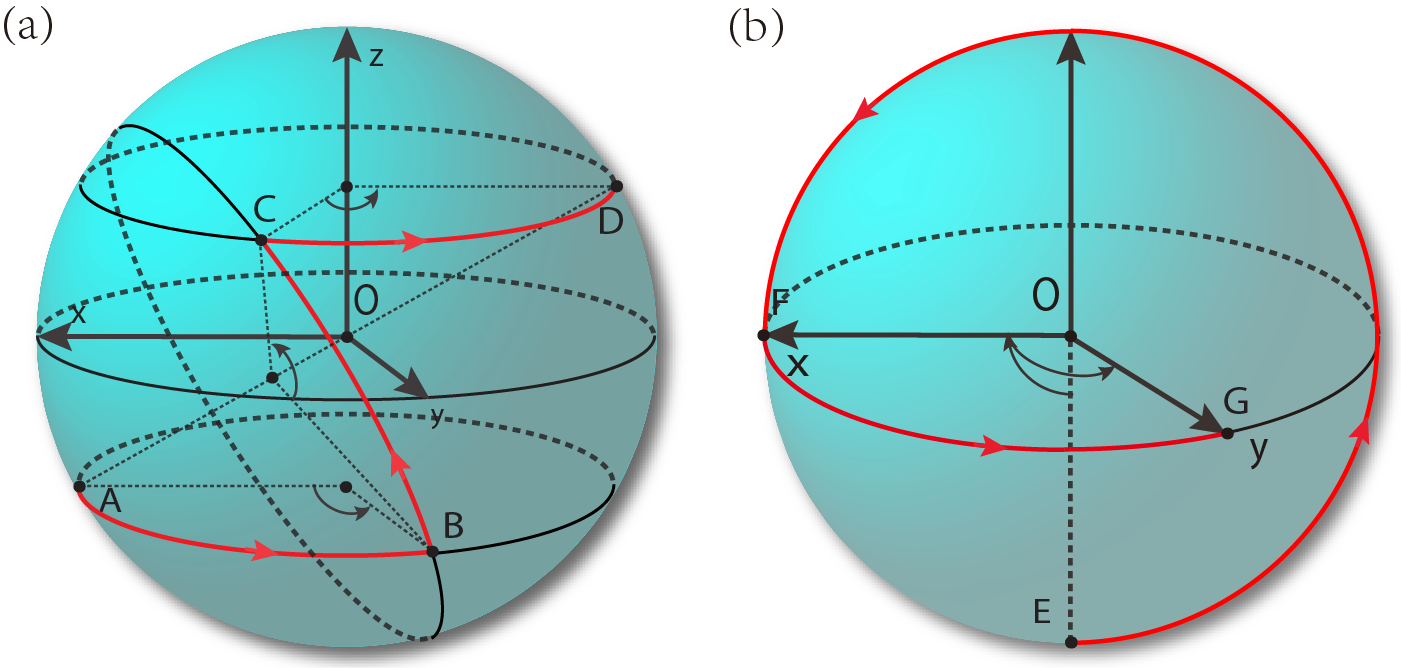}  \caption{Bloch sphere representation of evolutions for preparing the $W$ state. An arbitrary state in the subspace can be represented as a vector on the Bloch sphere: $|a\rangle=\left(\cos{\theta/2},e^{i\varphi}\sin{\theta/2}\right),$ where $\theta$ is the polar angle and $\varphi$ is the azimuthal angle. (a) Evolution $ZZ \left\{\left[\pi-\arccos{\left(1/3\right)}\right]/4\right\} -X\left[\arccos\left(1/3\right)\right]-ZZ \left\{\left [ \pi- \arccos {\left(1/3\right)}\right]/4\right\} $ for preparing the $W$ state in the subspace $\mathcal X_+$ for the three-qubit case. Two rotations around the $\hat z$ axis and a rotation around the axis $(\sqrt3/2,0,-1/2)$ rotate $ A(2\pi/3,0)= \left|x_1^+\right\rangle/2+ \sqrt3\left|x_2^+\right\rangle/2$ to the state $D(\pi/3,\pi)=\sqrt3\left|x_1^+\right\rangle/2-\left|x_2^+\right\rangle/2$. Here, $B$ and $C$ represent two intermediate states. Points $A$ and $D$ correspond to the points $A$ and $B$ in Fig. \ref{scheme} (a). (b) Evolution $Y\left(3\pi/8\right)-ZZ\left(\pi/12\right)$ for preparing the $W$ state in the subspace $\mathcal Y_-$ for the four-qubit case. A rotation around the $\hat y$ axis followed by a rotation around the $\hat z$ axis rotates  $E(\pi,\varphi)=\left|y_2^-\right\rangle$ to $G(\pi/2,\pi/2)=(i\left|y_1^-\right\rangle-\left|y_2^-\right\rangle)/\sqrt2$. Points $E$ and $G$ correspond to the points $C$ and $D$ in Fig. \ref{scheme}(b).}\label{3bit}
\end{figure}

For a three-spin case, following Eq. \eqref{steps}, we can obtain the $W$ state with the sequence:
\begin{equation}
\begin{split}
&Y(\pi/4)-ZZ\left(\left[\pi-\arccos{\left(\frac{1}{3}\right)}\right]/4\right)\\
&~~~-X\left(\arccos\left(\frac{1}{3}\right)/4\right)-ZZ\left(\left[\pi-\arccos{\left(\frac{1}{3}\right)}\right]/4\right)\\
&~~~~~~-Y(\pi/4).\\
\end{split}
\end{equation}
which can be represented on the Bloch sphere in Fig. \ref{3bit} (a). The detailed analysis is identical to that in \cite{gao}.

As for the four-spin case, according to Eq. \eqref{initial},
\begin{equation}\begin{split}
\left|\psi _{1}\right\rangle & =e^{-iH_y\pi/4}\left|0000\right\rangle\\
   & =\frac{\sqrt{2}}{4}\left|x^+_1\right\rangle
   +\frac{\sqrt{2}}{2}\left|x^+_2\right\rangle+\frac{\sqrt{6}}{4}\left|x^+_3\right\rangle.
\end{split}\end{equation}
We should find the sequence $U_x$ to rotate this state into the subspace $\mathcal Y_-$. Since $\left|x^+_2\right\rangle$ is the only common basis vector for the subspaces $\mathcal X_+$ and $\mathcal Y_-$,
suppose $U_x=X\left(\theta_1\right)-ZZ\left(\theta_2\right)-X\left(\theta_3\right)$, and let
\begin{equation}
e^{-iH_x^{\mathcal{X}_+}\theta_3}e^{-iH_{zz}^{\mathcal{X}_+}\theta_2}e^{-iH_{x}^{\mathcal{X}_+}\theta_1}\left|\psi_1\right\rangle=e^{i\delta}\left|x^+_2\right\rangle,
\label{dem}
\end{equation}
where we have $\theta_2=\pi/4$, $\theta_1=0$, and $\theta_3=\pi/16$ regarding simplicity.

According to Eq. \eqref{Weven},
$X\left(-\pi/4\right)\left| W\right\rangle=\left(i\left|y^-_1\right\rangle-\left|y^-_2\right\rangle\right)/\sqrt2$; we next need to find $U_y$, which rotates $\left|y_1^-\right\rangle$ to $(i\left|y^-_1\right\rangle-\left|y^-_2\right\rangle)/\sqrt2$. Since
\begin{equation}
H_y^{\mathcal{Y}_-}=\left(\begin{array}{cc}
0&-2i\\
2i&0
\end{array}\right)
\end{equation}
and
\begin{equation}
H_{zz}^{\mathcal{Y}_-}=\left(\begin{array}{cc}
6&0\\
0&0
\end{array}\right),
\end{equation}
we have
\begin{equation}
e^{-iH^{\mathcal Y_-}_{zz}\theta_1}e^{-iH^{\mathcal Y_-}_{y}\theta_2}\left(\begin{array}{c}0\\1\end{array}\right)
=\left(\begin{array}{c}-e^{-6i\theta_1}\sin{2\theta_2}\\ \cos{2\theta_2}\end{array}\right);
\end{equation}
when $\theta_1=\pi/12$ and $\theta_2=3\pi/8$, we obtain $\left(i\left|y^-_1\right\rangle-\left|y^-_2\right\rangle\right)/\sqrt2$. The process , as illustrated in Fig. \ref{3bit} (b), can be expressed as
\begin{equation}
E:\left(\pi,\varphi\right)\xrightarrow{{Y\left(3\pi/8\right)}}F:\left(\pi/2,0\right)
\xrightarrow{{ZZ\left(\pi/12\right)}}G:\left(\pi/2,\pi/2\right).
\label{54}
\end{equation}
Therefore, we achieve the target $W$ state with the following control sequence:
\begin{equation}
\begin{split}
Y\left(\pi/4\right)&-ZZ\left(\pi/4\right)-X\left(\pi/16\right)-Y\left(3\pi/8\right)\\
&-ZZ\left(\pi/12\right)-X\left(\pi/4\right)\rightarrow\left|W\right\rangle.\\
\end{split}
\end{equation}

\begin{figure}
\centering
\includegraphics[width=0.5\textwidth]{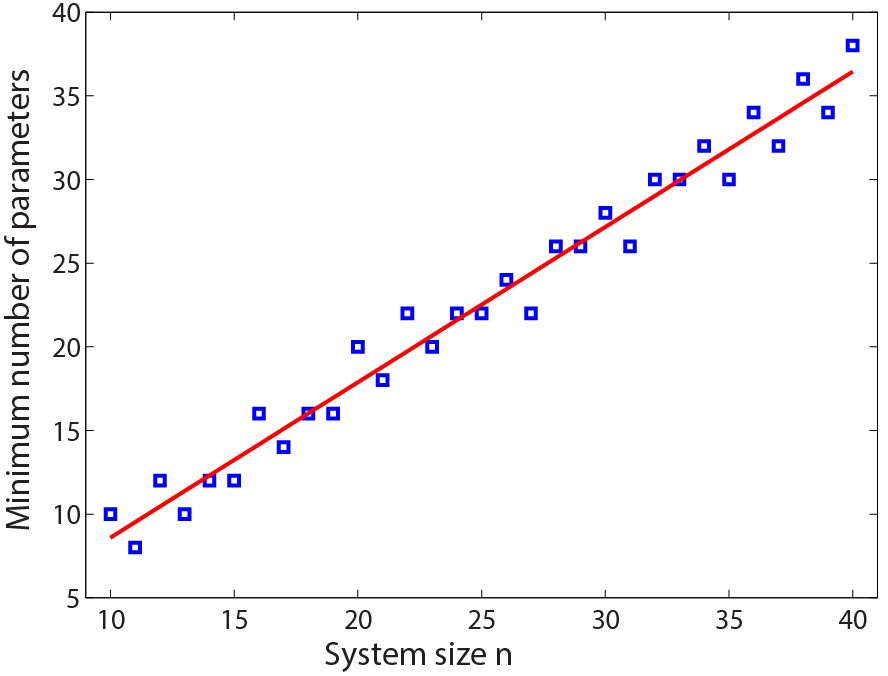}
\caption{Minimum number of parameters $\{ \tau_m, \beta _m^x,\beta _m^y\}$ to create $W$ states v.s. the system size $n$. The control sequence is $\{ZZ-X\}^i$  for odd-spin systems, and $\{ZZ-X\}^i-\{ZZ-Y\}^j$ for even-spin systems. Here, $i$ and $j$ denote the numbers of loops. The initial state is $Y(\pi/4)\left|00\cdots0\right\rangle$. In the numerical search, we set the final state to be $Y(-\pi/4)\left|W\right\rangle$ when $n$ is odd, and $X(-\pi/4)\left|W\right\rangle$ when $n$ is even.}
\label{rrrrsults}
\end{figure}

For large systems, it is difficult to find the analytic solutions. However, since the dimension of the subspaces increases linearly with the system size, with the numerical optimization algorithm it is possible to generate the $W$ states via the global-control sequence in principle. Applying the approach to a numerical search, one can find the solution in much larger systems (Fig. \ref{rrrrsults}). We tried 100 numerical searches with different initial parameters for each $n$ and different numbers of parameters.
Blue squares in Fig. \ref{rrrrsults} represent the minimum number of parameters in 100 sets of data to keep the fidelity above 0.999 by the numerical optimization algorithm. The red line represents the linear fitting. It is clear that the number of parameters increases almost linearly with the system size; that is, using only $O(n)$ operations we can prepare the $W$ states with a high fidelity in a long-range Ising model using only global controls.

%We remark here that the numerical solution is not unique, it is difficult to find the time-optimal solution.

\section{\label{sec:level6}Conclusion}

In summary, we presented a universal method to prepare the GHZ state and a standard procedure to prepare the $W$ state on a long-range Ising model using only global controls. Actually the solutions to preparing the GHZ state also obey the rules of the partition of subspaces described for the $W$ state: the GHZ state is within the subspace $\mathcal{X}_+$. One can follow a procedure similar to that used in finding solutions for the $W$ state to obtain different solutions for preparing the GHZ state if needed. Remarkably, the dimension of the irreducible subspace we employed increases only linearly with the size of the system due to the symmetry, so the scheme for preparing the $W$ state is also feasible in larger spin systems. In addition, it is also applicable to a wide range of physical implementations, and will contribute to quantum control for implementing quantum information processing in the future.

\section*{ACKNOWLEDGMENTS}
We thank Prof. D. D'Alessandro, I. Hinks, and H. Le for
helpful discussions. This work is supported by the National
Key Basic Research Program under Grant No. 2013CB921800
and No. 2014CB848700, the National Science Fund for
Distinguished Young Scholars under Grant No. 11425523,
the National Natural Science Foundation of China under
Grant No. 11375167 and No. 11661161018, the Strategic
Priority Research Program (B) of the CAS under Grant No.
XDB01030400.

\appendix*
\section{Controllability and irreducibility}
In this appendix, we give the matrix representations of the relevant Hamiltonians in the corresponding subspaces and proofs for controllability of the system and irreducibility of the subspaces $\mathcal X_\pm$ and $\mathcal Y_\pm$.

Let $H_x^s$, $H_y^s$ and $H_{zz}^s$ denote the matrix representations of Hamiltonians $H_y$, $H_y$ and $H_{zz}$ in the basis \eqref{basis}, respectively, and $\left\{a_{ij}\right\}$ and $\left\{b_{ij}\right\}$ denote the matrix elements of $H_x^s$ and $H_y^s$, respectively. All the elements of $H_x^s$ and $H_y^s$ are zero except those on the two lines next to the diagonal line:
\begin{equation}
\begin{cases}
 a_{k,k+1}=a_{k+1,k}=\sqrt{k(n-k+1)},\\
b_{k,k+1}=-b_{k+1,k}=-i\sqrt{k(n-k+1)}
\end{cases}
\label{ymatrix}
\end{equation}
for $k=1,\cdots,n$. The matrix representation of $H_{zz}$ is diagonal in this basis:
\begin{equation}
H_{zz}^s=\text{diag}\left(\lambda_1,\lambda_2,\cdots,\lambda_{n+1}\right),\label{zmatrix}
\end{equation}
where $\lambda_k=2\left(k-1-\frac{n}{2}\right)^2-\frac{n}{2}$ for $k=1,\cdots,n+1$.

From Fig. \ref{scheme} we know that it is only necessary to know the matrix representations of $H_x$ in the subspace $\mathcal X_+$ when $n$ is either even or odd, $H_y$ in the subspace $\mathcal Y_-$ when $n$ is even and $H_{zz}$. According to Eqs. \eqref{ymatrix} and \eqref{zmatrix}, we can write down the specific matrix representations of the Hamiltonians in the corresponding subspaces for $n \geqslant 2$.

\begin{figure}
\centering
\includegraphics[width=0.4\textwidth]{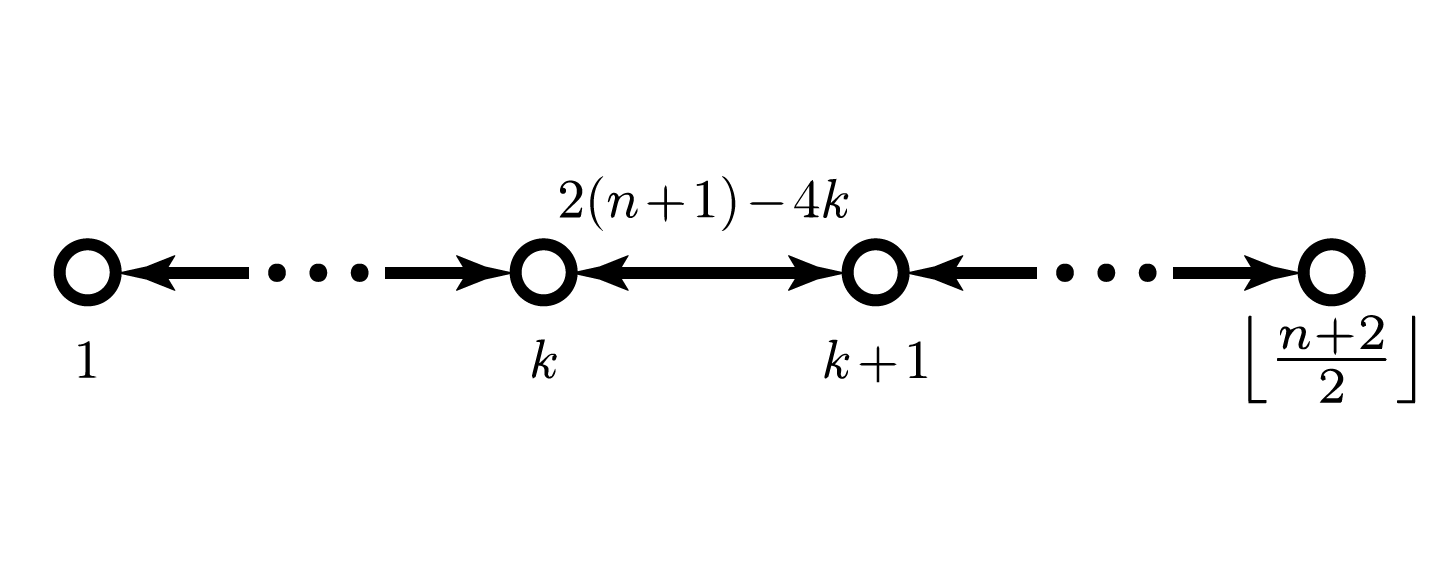}
\caption{Connectivity graph for $H_{zz}^{\mathcal X_+}$ and $H_{x}^{\mathcal X_+}$ in the subspace $\mathcal X_+$. Each vertex represents an eigenstate of $H_{zz}^{\mathcal X_+}$, and the label on the edge is the transition frequency. The graph is connected by the non-zero matrix elements of $H_x^{\mathcal X_+}$, and each transition is of a different frequency. The graph shows that the system is pure state controllable in the subspace $\mathcal X_+$}
\label{connect}
\end{figure}

When $n$ is even, the matrix elements of $H_x$, $\left\{ a_k^+\right\}$, in the subspace $\mathcal X_+$ are zero except those on the lines next to the diagonal line:
\begin{equation}
a_{k+1,k}^+= {a_{k,k+1}^+}=\begin{cases}\sqrt{k(n-k+1)} &  1\leqslant k\leqslant \frac{n}{2}-1,\\
\sqrt{n(n+2)/2} & k=\frac{n}{2}.\end{cases}
\label{x+even}
\end{equation}
When $n$ is odd, the matrix elements of $H_x$, $\left\{ \tilde{a}_k^+ \right\}$ (we use a tilde to denote the case when $n$ is odd), in the subspace $\mathcal X_+$ are zero except those on the lines next to the diagonal line and the last element at the bottom right corner:
\begin{equation}
\begin{cases} \tilde{a}_{k+1,k}^+={\tilde{a}_{k,k+1}^+}=\sqrt{k(n-k+1)} &  1\leqslant k\leqslant \frac{n-1}{2},\\
\tilde{a}_{\frac{n+1}{2},\frac{n+1}{2}}^+=\frac{n+1}{2} & k=\frac{n+1}{2}.\end{cases}
\label{x+odd}
\end{equation}
When $n$ is even, the matrix elements of $H_y$, $\left\{b_k^-\right\}$, in the subspace $\mathcal Y_-$ are zero except those on the lines next to the diagonal line:
\begin{equation}
b_{k+1,k}^-=-{b_{k,k+1}^-}=i\sqrt{k(n-k+1)},
\end{equation}
where $1\leqslant k\leqslant \frac{n}{2}-1$. Likewise, $H_{zz}$ is diagonal in $\mathcal{X}_\pm$ and $\mathcal{Y}_\pm$, and the matrix representations in these subspaces are:
\begin{equation}
H_{zz}^{\Xi}=\text{diag}\left(\lambda_1,\lambda_2,\cdots,\lambda_{\left\lfloor\frac{n+1}{2}\right\rfloor}\right),
\end{equation}
where ${\Xi}=\mathcal X_+,\mathcal X_-,\mathcal Y_+,\mathcal Y_-$.
All the diagonal elements are different from each other:
\begin{equation}
\lambda_k=2\left(k-1-\frac{n}{2}\right)^2-\frac{n}{2}~,~1 \leqslant k \leqslant \left\lfloor\frac{n+1}{2}\right\rfloor.
\label{diffe}
\end{equation}
When $n$ is even, $H_{zz}^{\mathcal X_+}$ and $H_{zz}^{\mathcal Y_+}$ include an additional diagonal element, $\lambda_{n/2+1}=-n/2$. Both the analytic solutions and the numerical search rely on the matrix representations of these Hamiltonians.

We are concerned with pure-state controllability of the system. One can prove this by looking at the connectivity graph where the vertices represent the eigenstates of $H_{zz}^{\mathcal X_+}$ and the edges connect vertices $j$ and $k$ if the element $H_x^{\mathcal X_+}(j,k)$ is nonzero \cite{shandeluo,turinici}. For example, in the subspace $\mathcal X_+$ (Fig. \ref{connect}), since the element $H_x^{\mathcal X_+}(k, k+1)$ is nonzero, it transfers magnitude between the $k$th and $(k+1)$th eigenstates of $H_{zz}^{\mathcal X_+}$.  Therefore, in the graph there is an edge connecting two adjacent vertices. Moreover, the difference between the adjacent eigenvalues $\lambda_k$ and $\lambda_{k+1}$ of $H_{zz}^{\mathcal X_+}$ is $2(n+1)-4k$, which means there is no degenerate transition. The same result holds for subspaces $\mathcal X_-$ and $\mathcal Y_\pm$. Therefore, the system is pure state controllable in the subspaces $\mathcal X_\pm$ and $\mathcal Y_\pm$ according to Theorem 3.7.1 in \cite{shandeluo}. This proof also guarantees the existence of the solutions in Eq. \eqref{steps}.

Next, we prove the subspaces $\mathcal X_\pm$ and $\mathcal Y_\pm$ are irreducible. Take $\mathcal X_+$ with odd $n$. In order to partition $\mathcal{X}_+$ further, one must find another operator $A$ such that $\left[H_x,A\right]=\left[H_{zz},A\right]=0$. Since the eigenstates of $H_{zz}$ are nondegenerate according to \eqref{diffe}, the matrix representation of $A$ must be diagonal:
\begin{equation}
A={\text {Diag}}\left(\lambda^A_1,\lambda^A_2,\cdots,\lambda^A_{\left\lfloor\frac{n+1}{2}\right\rfloor}\right).
\end{equation}
Since $\left[H_x,A\right]\left|s_k\right\rangle=\left(H_xA-AH_x\right)\left|s_k\right\rangle=0$, we can always obtain $\lambda^A_1=\lambda^A_2=\cdots=\lambda^A_{\left\lfloor\frac{n+1}{2}\right\rfloor}$ according to \eqref{x+even} and \eqref{x+odd}. Therefore, $A$ is a trivial matrix (a product of a constant and a unit matrix). Thus the subspace cannot be reduced any more. Similar results can be proved for the subspaces $\mathcal X_-$ and $\mathcal Y_\pm$ and for the case with even $n$.

\end{document}